\documentclass{mn2e}
\usepackage{graphicx}

\def\go{
\mathrel{\raise.3ex\hbox{$>$}\mkern-14mu\lower0.6ex\hbox{$\sim$}}
}
\def\lo{
\mathrel{\raise.3ex\hbox{$<$}\mkern-14mu\lower0.6ex\hbox{$\sim$}}
}
\def\simeq{
\mathrel{\raise.3ex\hbox{$\sim$}\mkern-14mu\lower0.4ex\hbox{$-$}}
}

\def\etal{{et al.\ }}

\def\msun{{\rm M_{\odot}}}

\begin{document}

\title[An X-ray Baldwin effect for the narrow Fe~K$\alpha$ lines observed in Active Galactic Nuclei]
{An X-ray Baldwin effect for the narrow Fe~K$\alpha$ lines observed in Active Galactic Nuclei}
\author[K.L. Page \etal]{K.L. Page$^{1}$, P.T. O'Brien$^{1}$, J.N. Reeves$^{2}$ and M.J.L. Turner$^{1}$\\
$^{1}$ X-Ray and Observational Astronomy Group, Department of Physics \& Astronomy,  
University of Leicester, LE1 7RH, UK\\
$^{2}$ Laboratory for High Energy Astrophysics, NASA Goddard Space Flight Center, Greenbelt, MD 20771, USA
}

\date{Received / Accepted}

\label{firstpage}

\maketitle

\begin{abstract}

The majority of Active Galactic Nuclei (AGN) observed by {\it XMM-Newton} reveal narrow Fe~K$\alpha$ lines at $\sim$~6.4~keV, due to emission from cold (neutral) material. There is an X-ray Baldwin effect in Type I AGN, in that the equivalent width of the line decreases with increasing luminosity, with weighted linear regression giving EW~$\propto$~L$^{-0.17~\pm~0.08}$ (Spearman Rank probability of $>$~99.9~per~cent). With current instrumental capabilities it is not possible to determine the precise origin for the narrow line, with both the Broad Line Region and putative molecular torus being possibilities. A possible explanation for the X-ray Baldwin effect is a decrease in covering factor of the material forming the fluorescence line. 

\end{abstract}


\section{Introduction}
\label{sec:intro}

As the number of AGN surveyed by {\it XMM-Newton} and {\it Chandra} increases, it is becoming apparent that the vast majority show evidence for a narrow (unresolved by {\it XMM}) line at $\sim$~6.4~keV, due to emission from neutral iron; recent papers describing such lines include Gondoin \etal (2003; NGC~3227), Page \etal (2003; Mrk~896), Pounds \etal (2003; Mrk~766), Petrucci \etal (2002; Mrk~841), Turner \etal (2002; NGC~3516), O'Brien \etal (2001; Mrk~359), Reeves \etal (2001; Mrk~205) and many others. In a number of cases (e.g., NGC~3783 -- Kaspi \etal 2002; NGC~5548 -- Yaqoob \etal 2001) the lines have actually been resolved by {\it Chandra}, with FWHM velocities typically $<$~5000~km~s$^{-1}$.

Iron K$\alpha$ emission was first identified as a common feature by {\it Ginga} (Pounds \etal 1990; Nandra \& Pounds 1994); observations by {\it ASCA} tended to find relatively broad profiles, although re-analysis of some of the data indicates that the lines {\it may} be narrower than originally measured (Lubinski \& Zdziarski 2001). There is some disagreement over these results, however, with Yaqoob \etal (2002) stating that the {\it ASCA} calibration changes have a negligible effect on the line profiles. Very few broad lines have been found in {\it XMM} data to date, examples being MCG~$-$6$-$30$-$15 (Fabian \etal 2002), Mrk~205 (Reeves \etal 2001) and Mrk~509 (Pounds \etal 2001). However, not all these line profiles are the same, with MCG~$-$6$-$30$-$15 showing a strongly asymmetric line, presumably due to the strong gravitational forces and high velocity in the inner accretion disc. Mrk~205, however, is only well-fitted by a relativistic disc-line model if the disc is strongly ionised, since the broad emission peaks at $\sim$~6.7~keV. This is not the only conceivable explanation, as Reeves \etal (2001) discuss: the emission could come from a spherical distribution of clouds, rather than the planar accretion disc; alternatively, the broad line may actually consist of a blend of different ionisation narrow lines. The broad line in Mrk~509 is also apparently ionised.

The narrow emission lines observed by {\it XMM-Newton} and {\it Chandra} are interpreted as Fe~K fluorescence from cold (neutral) matter far from the inner accretion disc. Suggestions for the origin of the narrow line include the putative molecular torus, the Broad Line Region (BLR) or the very outer-most reaches of the accretion disc.

The Baldwin effect is well known for optical/UV emission lines, with Baldwin (1977) first reporting that the EW of the C~{\sc iv} ($\lambda$~=~1549~\AA)  line decreased with increasing UV luminosity. Since then, significant anti-correlations have been found between the luminosity and various other ions, such as N~{\sc v}, He~{\sc ii}, C~{\sc iii}], Mg~{\sc ii} and Ly$\alpha$ (e.g., Tytler \& Fan 1992; Zamorani \etal 1992; Green, Forster \& Kuraszkiewicz 2001; Dietrich \etal 2002), although the strengths of these correlations are still unclear. 

It should be noted that, although the Baldwin effect is generally accepted to be an anti-correlation between equivalent width and luminosity, Green \etal (2001) claim that EW is actually more strongly correlated with redshift than luminosity for their data. However, Croom \etal  (2002) find that, for 12 of the 14 lines tested, the stronger correlation is with absolute magnitude rather than redshift.

Iwasawa \& Taniguchi (1993) reported an X-ray Baldwin effect in the Fe~K lines found in {\it Ginga} observations of AGN. They find a strong relationship for their Seyfert sample, but were unable to conclude that it holds for QSOs, due to poor constraints; there is also a Baldwin effect for the broad iron lines found in {\it ASCA} data (Nandra \etal 1997). Such broad lines are thought to be produced through fluorescence of the accretion disc itself. Nandra \etal suggest, based on an earlier paper (Nandra \etal 1995), that this Baldwin effect is due to the presence of an ionised skin on the accretion disc, with the degree of ionisation increasing with luminosity; see also Nayakshin (2000a,b). Nandra \etal (1997) also find that the narrow line core drops in intensity as the luminosity increases, but conclude that the entire Baldwin effect in their data can be attributed to the broader line alone.

In this paper we show, and attempt to explain, a Baldwin effect for the narrow Fe~K$\alpha$ lines measured in {\it XMM} data, for a sample including both Seyfert galaxies and QSOs.

\section{XMM-Newton Observations}
\label{sec:xmmobs}

This sample consists of 53 type I AGN (Table~\ref{objects}), these being mainly a combination of our own proprietary targets and public observations obtained from the online {\it XMM} Science Archive\footnote{http://xmm.vilspa.esa.es/external/xmm\_data\_acc/xsa/index.shtml}. A literature search also revealed four more objects for which the relevant data had been published (NGC~5506 -- Matt \etal 2001; NGC~3516 -- Turner \etal 2002; 1H~0707$-$495 -- Boller \etal 2002; Ton~S180 -- Vaughan \etal 2002 and  Vaughan 2003, private communication).

\begin{table}
\begin{center}
\caption{The type I AGN included in this sample, ordered by redshift within the radio-quiet and radio-loud groups. The luminosities were calculated for the 2--10~keV rest frame and 90~per~cent errors/upper limits are given for the rest-frame equivalent widths. $^{a}$ Narrow Line Seyfert 1 galaxies; $^{b}$ Broad Line Seyfert 1 galaxies; $^{c}$ QSOs} 
\label{objects}
\begin{tabular}{p{2.2truecm}p{0.5truecm}p{1.0truecm}p{1.3truecm}p{1.2truecm}}
\hline
object & type & redshift & log[lum. & EW\\
 & & & (erg s$^{-1}$)] & (eV)\\
\hline
NGC~4151$^{b}$ & RQ & 0.003 & 42.27 & 187~$\pm$~3\\
NGC~5506$^{b}$ & RQ & 0.006 & 42.83 & 70~$\pm$~20\\
MCG~$-$6$-$30$-$15$^{b}$ & RQ & 0.008 & 42.90 & 52~$\pm$~10\\
NGC~3516$^{b}$ & RQ & 0.009 & 42.39 & 196~$\pm$~22\\
NGC~4593$^{b}$ & RQ & 0.009 & 43.07 & 98~$\pm$~21\\
Mrk~766$^{a}$ & RQ & 0.013 & 43.16 & 45~$\pm$~35\\
IC~4329a$^{b}$ & RQ & 0.016 & 44.77 & 30~$\pm$~12\\
Mrk~359$^{a}$ & RQ & 0.017 & 42.49 & 220~$\pm$~74\\
Mrk~1044$^{a}$ & RQ & 0.017 & 42.55 & 186~$\pm$~61\\
NGC~5548$^{b}$ & RQ & 0.017 & 43.39 & 59~$\pm$~6\\
Mrk~335$^{a}$ & RQ & 0.026 & 43.27 & $<$~54\\
Mrk~896$^{a}$ & RQ & 0.026 & 42.70 & 180~$\pm$~87\\
Mrk~493$^{a}$ & RQ & 0.031 & 43.20 & $<$~101\\
Mrk~509$^{b}$ & RQ & 0.034 & 44.68 & 85~$\pm$~57\\
Mrk~841$^{b}$ & RQ & 0.036 & 43.89 & $<$~83\\
1H~0707$-$495$^{a}$ & RQ & 0.041 & 42.23 & $<$~90\\
ESO~198$-$G24$^{b}$ & RQ & 0.046 & 43.72 & 104~$\pm$~49\\
Fairall 9$^{b}$ & RQ & 0.047 & 43.97 & 139~$\pm$~26\\
Mrk~926$^{b}$ & RQ & 0.047 & 44.14 & $<$~61 \\
Ton~S180$^{a}$ & RQ & 0.062 & 43.62 & $<$~64\\
MR~2251$-$178$^{c}$ & RQ & 0.064 & 44.46 & $<$~74\\
PG~0844+349$^{c}$ & RQ & 0.064 & 43.88 & $<$~172\\
Mrk~304$^{c}$ & RQ & 0.066 & 43.82 & $<$~115\\
Mrk~205$^{c}$ & RQ & 0.071 & 43.95 & 60~$\pm$~25\\
PG~1211+143$^{c}$ & RQ & 0.081 & 43.63 & 37~$\pm$~24\\
HE~1029$-$1401$^{c}$ & RQ & 0.086 & 44.54 & $<$~105\\
Mrk~1383$^{c}$ & RQ & 0.086 & 44.10 & 77~$\pm$~46\\
PG~0804+761$^{c}$ & RQ & 0.100 & 44.32 & $<$~101\\
1H~0419$-$577$^{b}$ & RQ & 0.104 & 44.56 & $<$~85 \\
Mrk~876$^{c}$ & RQ & 0.129 & 44.19 & 96~$\pm$~59\\
PG~1626+554$^{c}$ & RQ & 0.133 & 44.38 & $<$~281\\
PG~1114+445$^{c}$ & RQ & 0.144 & 44.27 & 120~$\pm$~17\\
Q~0056$-$363$^{c}$ & RQ & 0.162 & 44.23 & $<$~159\\
PG~1048+342$^{c}$ & RQ & 0.167 & 44.21 & $<$~191\\
PDS~456$^{c}$ & RQ & 0.184 & 44.77 & $<$~18\\
Q~0144$-$3938$^{c}$ & RQ & 0.244 & 44.34 & 152~$\pm$~50\\
UM~269$^{c}$ & RQ & 0.308 & 44.46 & $<$~99\\
PG~1634+706$^{c}$ & RQ &1.334 & 46.11 & $<$~63\\
PB~05062$^{c}$ & RQ & 1.77 & 46.85 & $<$~61\\

NGC~3227$^{c}$ & RL & 0.004 & 41.37 & 191~$\pm$~23\\
PKS~1637$-$77$^{c}$ & RL & 0.043 & 43.73 & $<$~95\\
3~Zw~2$^{c}$ & RL & 0.089 & 44.34 & $<$~84\\
PKS~0558$-$504$^{c}$ & RL & 0.137 & 44.63 & $<$~11\\
3C~273$^{c}$ & RL & 0.158 & 45.72 & $<$~9\\
B2~1028+31$^{c}$ & RL & 0.178 & 44.26 & 88~$\pm$~41\\
B2~1721+34$^{c}$ & RL & 0.206 & 44.90 & 63~$\pm$~34\\
B2~1128+31$^{c}$ & RL & 0.289 & 44.72 & 50~$\pm$~39\\
S5~0836+71$^{c}$ & RL & 2.172 & 47.69 & 16~$\pm$~6\\
PKS~2149$-$30$^{c}$ & RL & 2.345 & 47.03 & $<$~40\\
PKS~0438$-$43$^{c}$ & RL & 2.852 & 46.91 & $<$~26\\
PKS~0537$-$286$^{c}$ & RL & 3.104 & 47.43 & $<$~82\\
PKS~2126$-$15$^{c}$ & RL & 3.268 & 47.45 & $<$~17\\
S5~0014+81$^{c}$ & RL & 3.366 & 47.20 & $<$~17\\
\hline
\end{tabular}
\end{center}
\end{table}

For the spectra analysed here, the PN data only were used, since the instrument is more sensitive than the MOS cameras at higher energies. For each spectrum, a simple power-law was fitted over the rest frame band of 2--10~keV. If there was evidence for any broadened emission, then a broad Gaussian line was included; see Table~\ref{broad}. A narrow ($\sigma$~=~10~eV) line was next added and the equivalent width (EW) measured. If the reduction in $\chi^{2}$ was less than 99~per~cent significant, the 90~per~cent upper limit was taken for the equivalent width. For the vast majority of the spectra, the energy of such a line was very close to 6.4~keV, and consistent with being neutral within the errors. For four of the higher luminosity objects, the line energy could not be constrained, with errors much bigger than the actual value. Six of the AGN tended towards ionised lines, although, when taking account of the errors, only two were inconsistent with a neutral origin, these being PKS~0558$-$504 and PKS~2126$-$15. For these objects with unconstrained or ionised line energies, the  centre of the line was fixed at 6.4~keV. A neutral reflection component ({\it pexrav} in {\sc xspec}; Magdziarz \& Zdziarski 1995) was also tried for each spectrum. Only a small number were improved by this addition and, of these reflection parameters, the vast majority were unconstrained. The two objects which gave significant, constrained values were NGC~4151 (R~=~0.52~$\pm$~0.05) and MCG~$-$6$-$30$-$15 (R~=~1.81~$\pm$~0.52); R~=~$\Omega$/2$\pi$, where $\Omega$ is the solid angle subtended by the reflecting matter. Such a reflection component is evidence for scattering off cool, optically thick matter (i.e., the torus) and will be discussed in more detail later. Throughout this paper, H$_{0}$ is taken to be 70~km~s$^{-1}$~Mpc$^{-1}$, q$_{0}$~=~0.5 and $\Lambda$~=~0.7.

\begin{table}
\begin{center}
\caption{The objects in the sample for which broad lines were statistically required. $^{a}$ The broad line fit to MCG~$-$6$-$30$-$15 was based on that by Wilms \etal (2001), which used a relativistic {\sc laor} line, rather than a broad Gaussian. Both the energy and index ($\beta$~=~4.6) were frozen at their best fit values from Wilms \etal} 
\label{broad}
\begin{tabular}{p{2.0truecm}p{1.5truecm}p{1.8truecm}p{1.5truecm}}
\hline
object & line energy & intrinsic & EW \\
 & (keV) & width (keV) & (eV)\\
\hline
MCG~$-$6$-$30$-$15 &  6.95 & see note a & 590~$\pm$~62\\
Mrk~766 & 6.78~$\pm$~0.19 & 0.230~$\pm$~0.150 & 70~$\pm$~30\\
Mrk~335 & 6.72~$\pm$~0.14 & 0.439~$\pm$~0.129 & 175~$\pm$~50\\
Mrk~509 & 6.84~$\pm$~0.14 & 0.149~$\pm$~0.086 & 28~$\pm$~20\\
Fairall~9 & 6.91~$\pm$~0.141 & 0.464~$\pm$~0.155 &  206~$\pm$~90\\
Mrk~926 & 6.24~$\pm$~0.15 & 0.189~$\pm$~0.117 & 91~$\pm$~35\\
PG~0844+349 & 6.47~$\pm$~0.09 & 0.345~$\pm$~0.117 & 334~$\pm$~130\\
Mrk~205 & 6.86~$\pm$~0.11 & 0.195~$\pm$~0.151 & 122~$\pm$~80\\
PG~1211+143 & 6.50~$\pm$~0.20 & 0.120~$\pm$~0.065 & 240~$\pm$~76\\
PG~0804+761 & 7.00~$\pm$~0.36 & 0.385~$\pm$~0.210 & 1050~$\pm$~500\\
Q~0056$-$363 & 6.32~$\pm$~0.07 & 0.218~$\pm$~0.079 & 260~$\pm$~116\\
B2~1028+31 & 6.56~$\pm$~0.09 & 0.206~$\pm$~0.101 & 131~$\pm$~110\\
B2~1128+31 & 6.41~$\pm$~0.33 & 0.693~$\pm$~0.437 & 253~$\pm$~185\\
\hline
\end{tabular}
\end{center}
\end{table}

\begin{figure}
\begin{center}
\includegraphics[clip,width=0.7\linewidth,angle=-90]{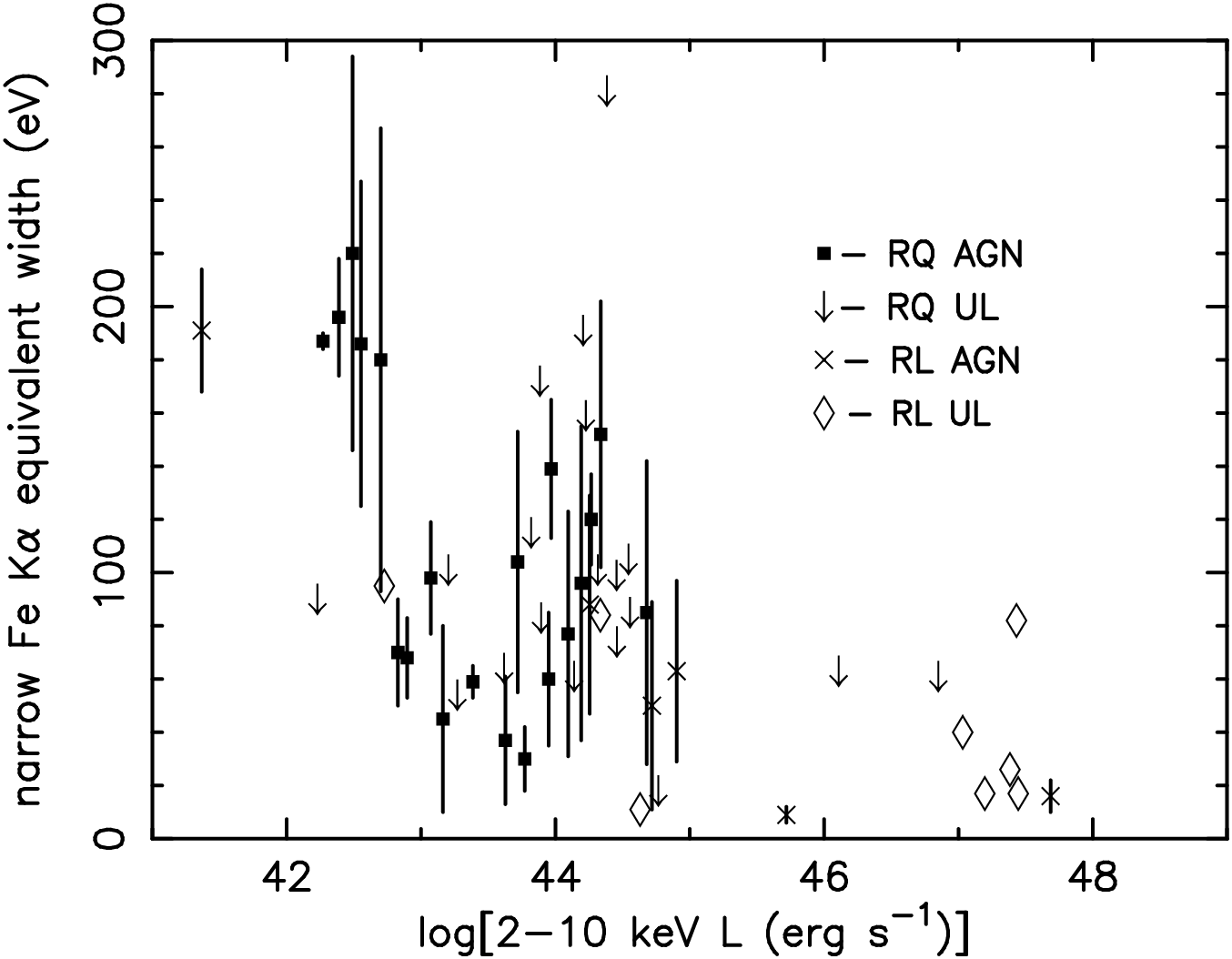}
\includegraphics[clip,width=0.7\linewidth,angle=-90]{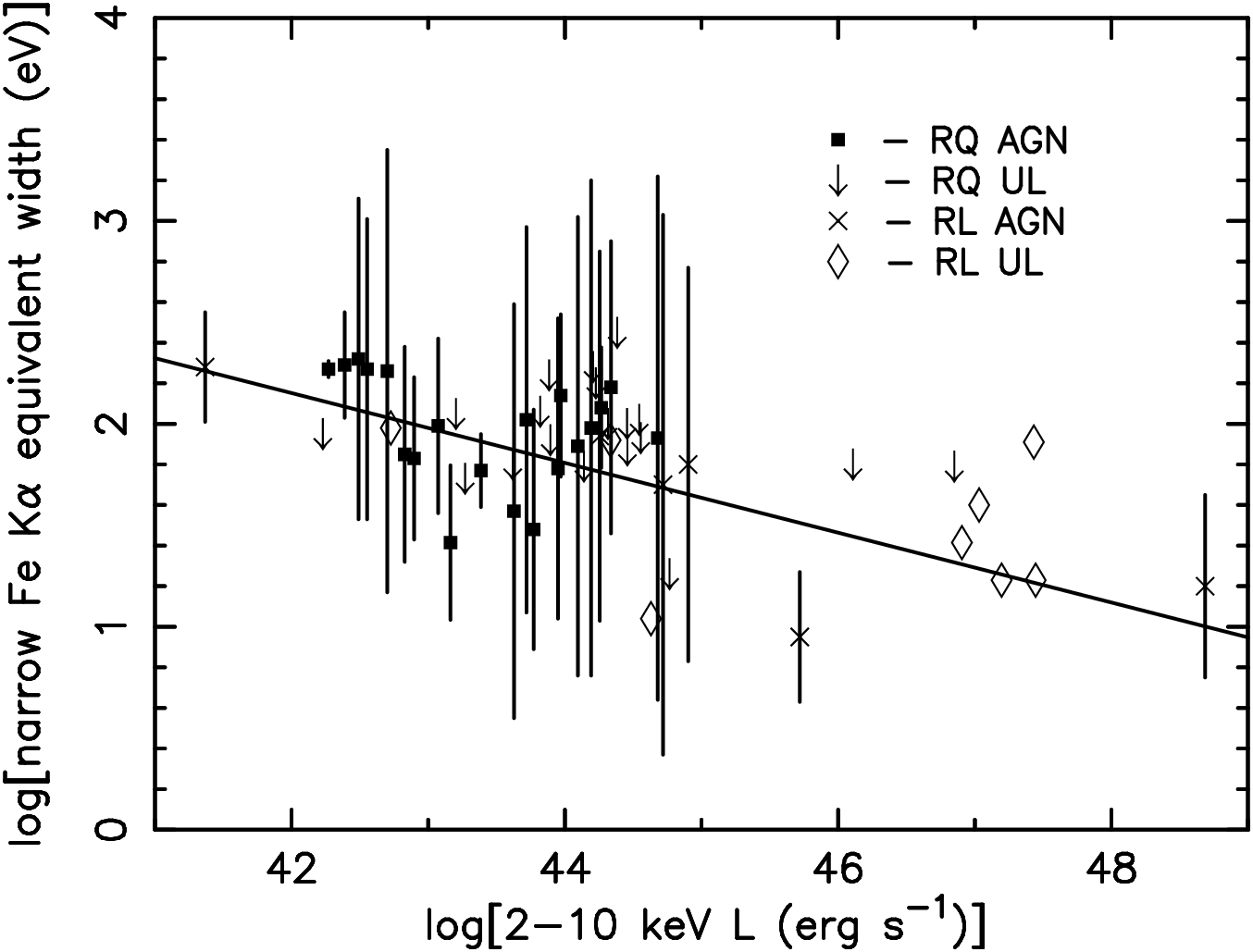}
\caption{The decrease in EW of the narrow, neutral iron line with luminosity. Statistically significant measurements are indicated by squares and crosses for radio-quiet and radio-loud objects respectively. Arrows and diamonds signify the corresponding upper limits. The line in the bottom plot shows a power-law fit to the data: EW~$\propto$~L$^{-0.17}$.}
\label{EW-lum}
\end{center}
\end{figure}

\begin{figure}
\begin{center}
\includegraphics[clip,width=0.7\linewidth,angle=-90]{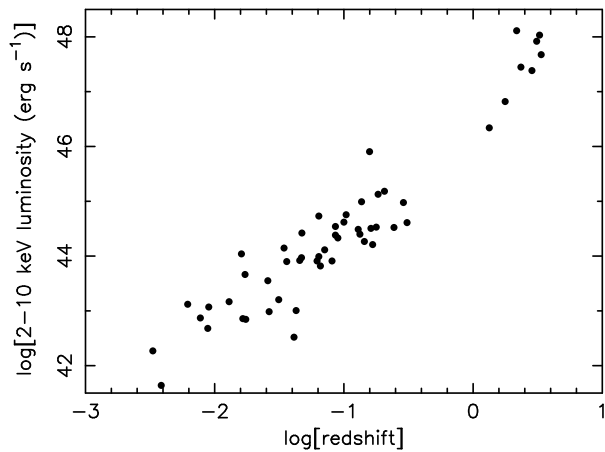}
\caption{The correlation between luminosity and redshift for the sample of AGN.}
\label{lum-z}
\end{center}
\end{figure}

Figure~\ref{EW-lum} plots the rest-frame EW against the de-reddened 2--10~keV luminosity. It can clearly be seen that there is a decrease in the EW as the luminosity increases -- the `X-ray Baldwin effect'. The {\sc asurv} (Astronomy Survival Analysis; Feigelson \& Nelson 1985) package can be used in the presence of censored (upper limit) data. This allows the Spearman Rank (SR) statistic to be applied to the complete dataset, and gives an anti-correlation between the EW and luminosity of $>$~99.98~per~cent ($\sim$~98.5~per~cent if the upper limits are dropped). Due to selection effects, luminosity and redshift are very strongly correlated, as shown in Figure~\ref{lum-z}. Hence, it is often difficult to determine whether the correlation in question is with luminosity or redshift. Using the simple Spearman Rank, a weaker correlation (99~per~cent; 79.5~per~cent if the upper limits are excluded) was found between the line strength and the redshift. An alternative method for checking involves the Partial Spearman Rank, which gives an indication as to which of the two relationships is stronger. In this case, agreeing with the simple Spearman Rank, the EW-luminosity correlation appears to be the dominant relationship.


Weighted linear regression can be used to find the slope of the best-fit line to the log-log plot of EW against luminosity -- that is, the power-law fit. This was performed using two different methods. Firstly the upper limit measurements were completely removed; this gave a value of $-$0.17~$\pm$~0.08. The second method used the linear regression option within {\sc asurv}; although upper limit values are accounted for, this method does not include the errors on the actual measurements. Using this method, the slope is very similar, with a value of $-$0.18~$\pm$~0.04.


As a further check to determine whether the Baldwin effect could be due to evolution, the objects at higher redshifts were progressively removed from the sample. It was found, however, that even considering only those AGN at z~$<$~0.1, there still remained an inverse correlation between EW and luminosity which was consistent in magnitude (slope of 0.15~$\pm$~0.08 using {\sc asurv}) with that found for the complete sample. This implies that the decrease in line strength is predominantly a luminosity-dependent effect.

\begin{figure}
\begin{center}
\includegraphics[clip,width=0.7\linewidth,angle=-90]{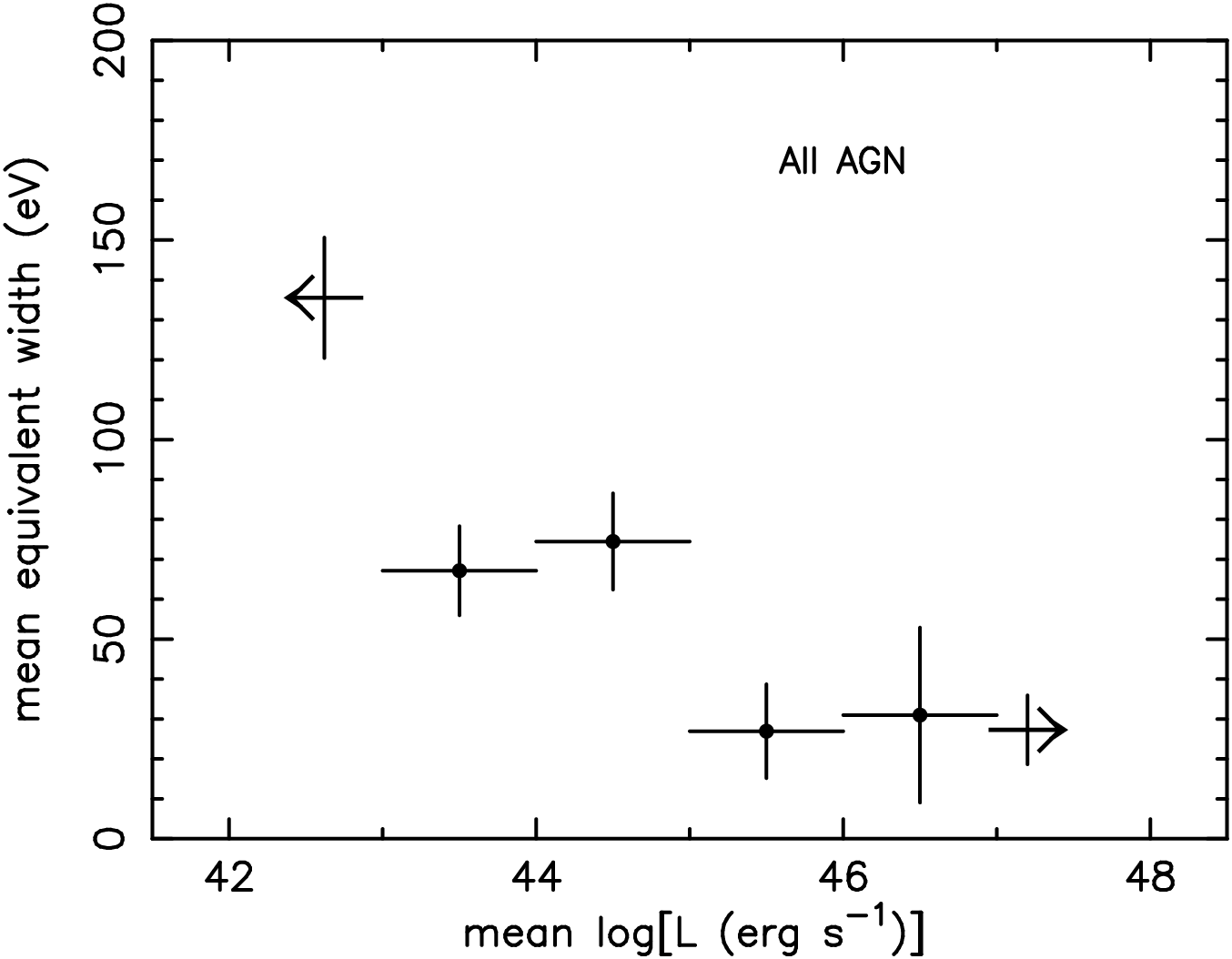}\vspace*{0.3cm}
\includegraphics[clip,width=0.7\linewidth,angle=-90]{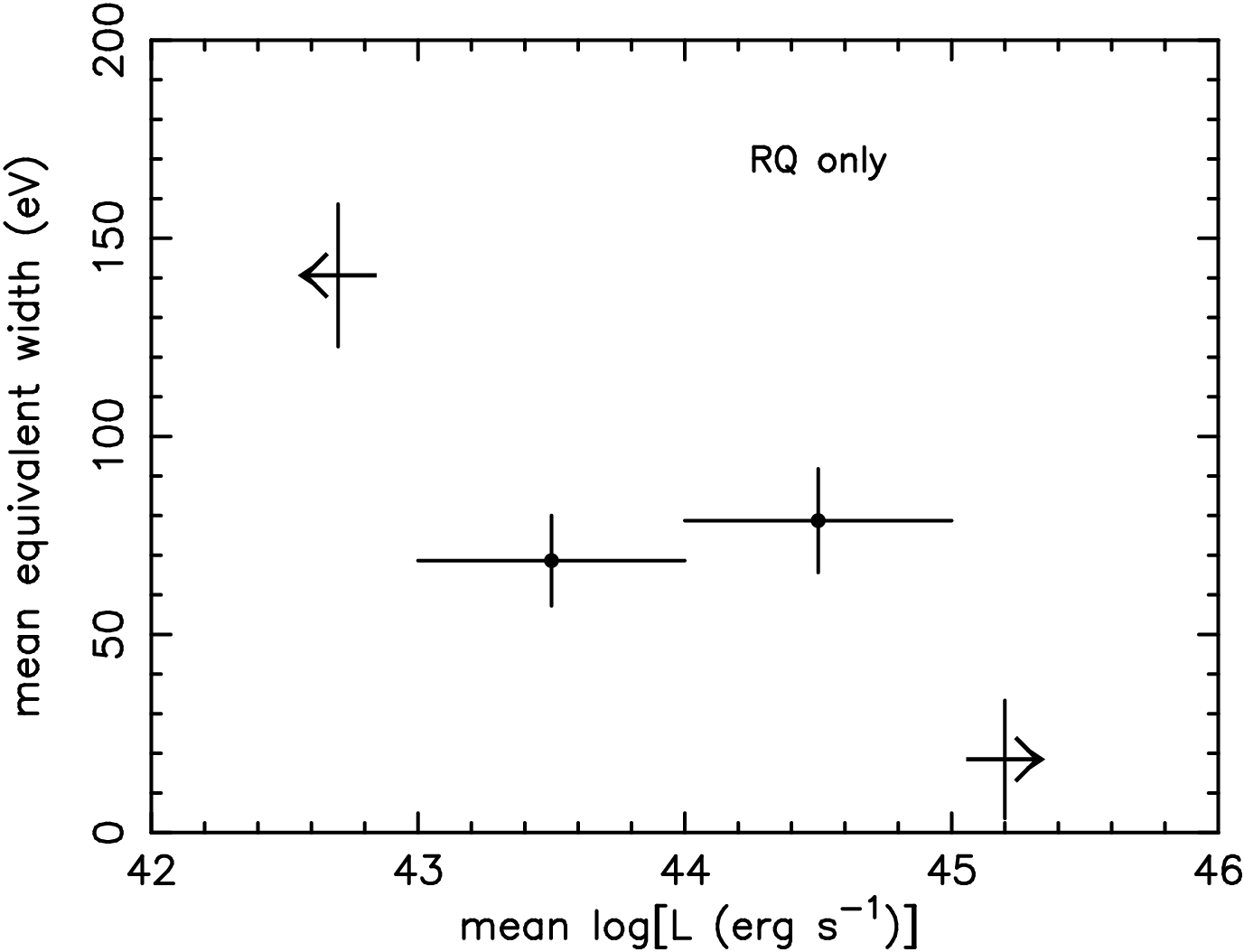}
\caption{Averaging the iron line equivalent widths within luminosity bins clearly shows the X-ray Baldwin effect, even when excluding the radio-loud objects (bottom plot).}
\label{bins}
\end{center}
\end{figure}

There is a possible complication in that the highest luminosity AGN tend to be radio-loud, since the average X-ray emission from radio-quiet AGN is about three times lower than that from radio-loud objects (Zamorani \etal 1981; Worrall \etal 1987). The underlying worry, therefore, is that the EW is very low at the highest luminosities because of dilution through beaming effects. However, as Fig.~\ref{EW-lum} shows, there are lower-luminosity radio-loud AGN in this sample which have correspondingly higher EWs. Conversely, PB~5062 and PG~1634+706 -- radio-quiet AGN -- at luminosities of $\sim$~10$^{46}$~erg~s$^{-1}$ have low upper limits of EW~$<$~63~eV for their lines. Hence, the anti-correlation observed is not simply due to beaming. To show the overall effect more clearly, Figure~\ref{bins} was plotted. This shows the mean EW for a number of luminosity bins considering all the objects (top plot) and just the radio-quiet ones (bottom). In both plots, the decrease in EW is obvious. To produce Figure~\ref{bins}, upper limits were taken to be half of the value, together with an equally sized error bar; a similar decrease in EW is also observed if the upper limits are dropped, although the luminosity range covered is lower.


\begin{figure}
\begin{center}
\includegraphics[clip,width=0.7\linewidth,angle=-90]{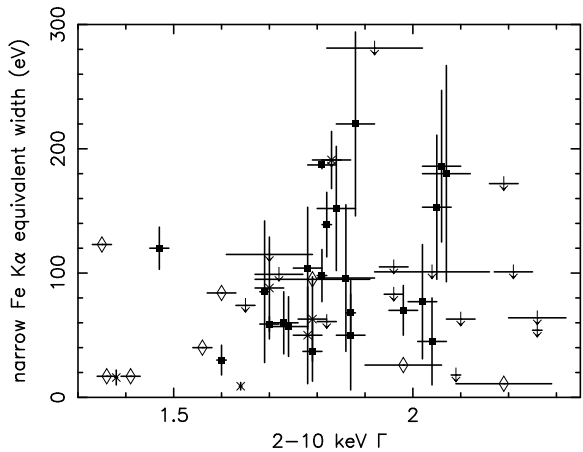}
\caption{The equivalent width of the Fe K$\alpha$ line against the rest frame 2--10~keV photon index. Symbols as in Figure~\ref{EW-lum}.}
\label{ew-gamma}
\end{center}
\end{figure}

Figure~\ref{ew-gamma} plots the EW against the power-law slope measured over the 2--10~keV rest-frame energy band. Since flatter slopes are often taken to be indicative of reflection, it might be expected that they would correspond to stronger emission lines (since the fluorescence line may come about through reflection processes), although the 2--10~keV band is not very sensitive to reflection components. However, this is clearly not the case; if anything, there is a slight trend in the opposite direction, with some of the steeper slopes showing the highest EWs. 


\section{Discussion}
\label{sec:disc}

Suggestions for the origin of the neutral Fe~K$\alpha$ line include
the putative molecular torus and the BLR; these theories will be
investigated in the following sections, considering both the
feasibility of the line production and possible explanations for the
X-ray Baldwin effect.

\subsection{The broad-line region}
\label{blr}

Yaqoob \etal (2001) give the following equation for the EW of the
narrow line in AGN, for low optical depths:

\begin{displaymath}
EW_{Fe K\alpha} = 42\left(\frac{f_{c}}{0.35}\right)\left(\frac{\omega_{K}}{0.34}\right)\left(\frac{A_{Fe}}{4.68 \times 10^{-5}}\right)\left(\frac{N_{H}}{10^{23}}\right) 
\end{displaymath}
\begin{displaymath}
\;\;\;\;\;\;\;\;\;\;\;\;\;\;\;\;\;\;\;\times \left(\frac{3.2}{\Gamma + 1.646}\right)\left(\frac{E_{K}}{7.11}\right)\left(\frac{E_{K\alpha}}{E_{K}}\right)^{\Gamma} eV
\end{displaymath}

\noindent where E$_{K}$ is the energy of Fe K edge (7.11~keV for neutral
iron) and E$_{K\alpha}$ the central energy of the line, both in units of keV. f$_{c}$ is the
fraction of the sky covered by BLR clouds; A$_{Fe}$, the abundance of
iron relative to hydrogen, where 4.68~$\times$~10$^{-5}$ is the Solar
value (Anders \& Grevesse 1989); N$_{H}$, the column density of the clouds in cm$^{-2}$ and $\omega_{K}$, the
fluorescence yield (=~0.34 for neutral Fe).

Based on a particular BLR model, Yaqoob \etal find that the strength
of the Fe~K$\alpha$ line in NGC~5548 is likely to be too large to come
simply from the BLR; the predicted EW is 36~eV, whereas the measured
value is almost four times larger, at 133~eV. They suggest that these
differences could be explained by a decline in the continuum level
shortly before the observation; in this event, the line would not have
had time to respond to the smaller continuum level and, so, would
appear unusually strong. 

Whatever the explantion for NGC~5548, if the BLR origin is correct, the reason  why most of the objects in Table~\ref{objects} have relatively high
EW values compared to those predicted by the above equation must be explained. It must also be determined 
whether a variation in the parameters could explain the observed EW-luminosity correlation. The iron K line is produced by a different
physical process (fluoresence) from the optical/UV emission lines
(photoionisation). Nevertheless, changes in covering factor or
abundance, for example, might be reflected in the BLR optical/UV lines,
so it is briefly examined whether such correlations exist.

\begin{figure}
\begin{center}
\includegraphics[clip,width=0.7\linewidth,angle=-90]{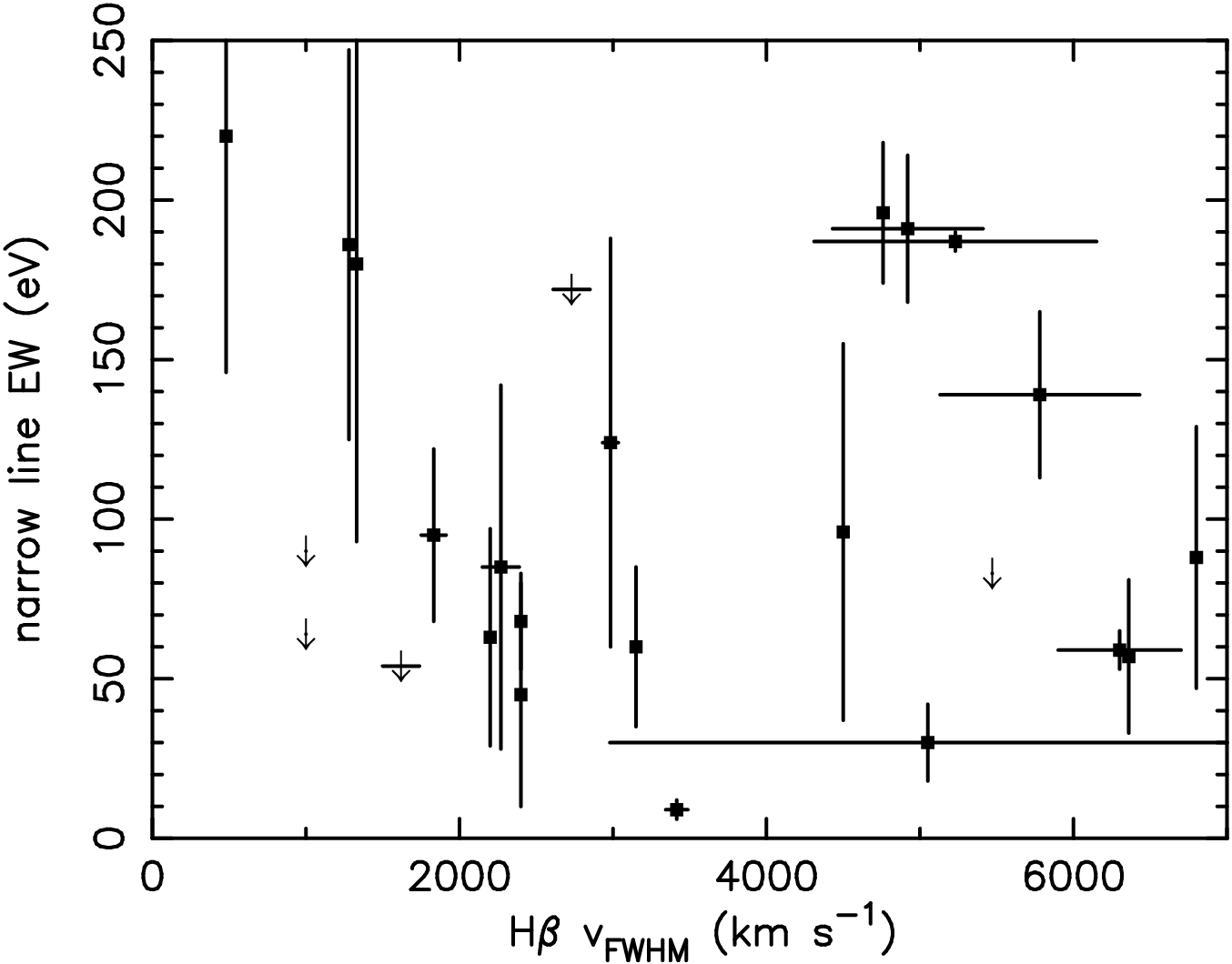}\vspace*{0.3cm}
\includegraphics[clip,width=0.7\linewidth,angle=-90]{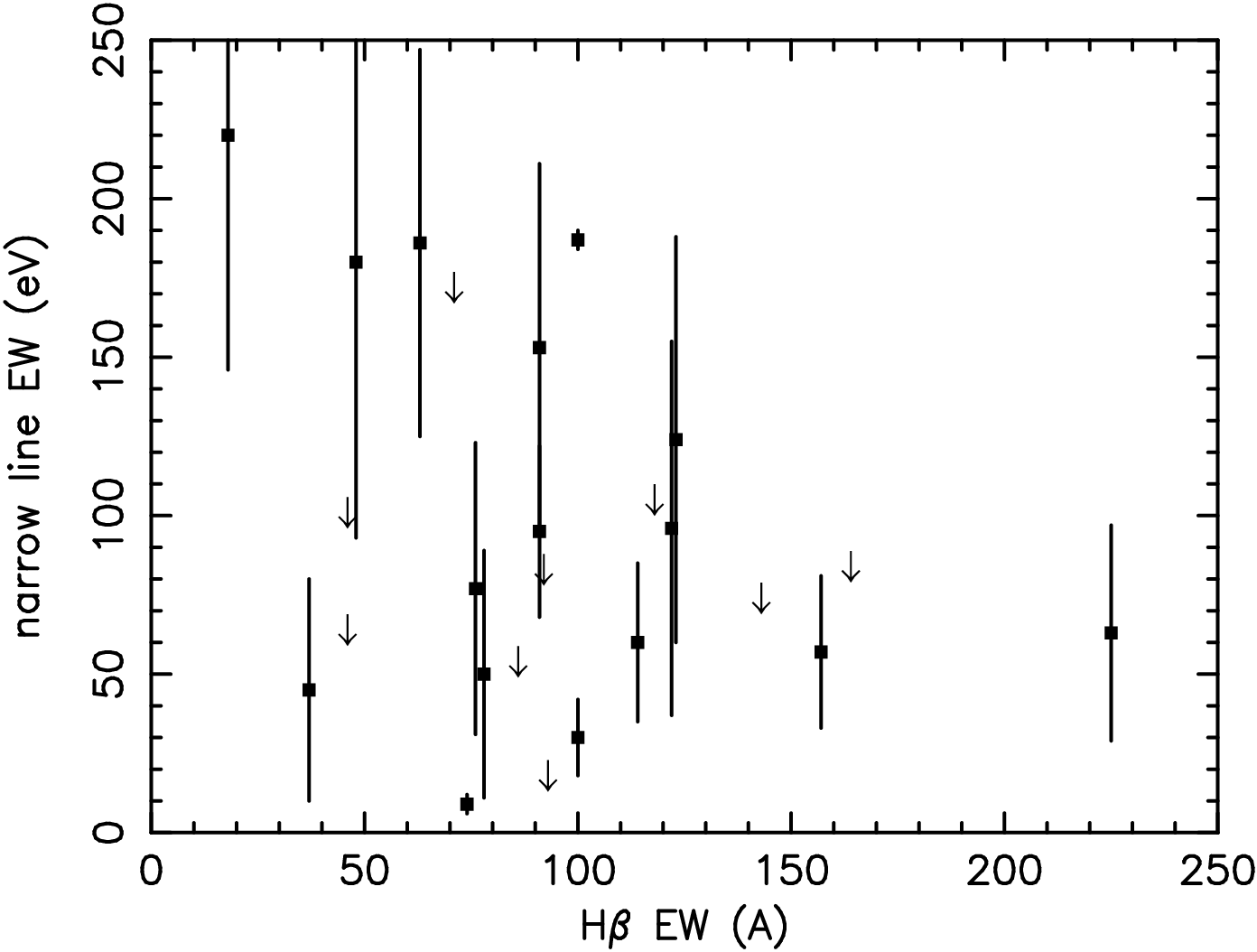}
\caption{A comparison of the equivalent width of the narrow Fe~K$\alpha$ line and the FWHM and EW of H$\beta$.}
\label{hbetafwhm}
\end{center}
\end{figure}

One possibility is that the covering fraction of the BLR clouds is
systematically underestimated in the above formula and decreases with
increasing luminosity. There is no independent evidence for very high
BLR covering factors, however, and the observed EW of optical/UV lines can be
reproduced using modest values (few 10s of percent). For the
AGN sample used here, no correlation is found between the EW of the
narrow iron line and those of C{\sc iv} $\lambda$1549 and Ly$\alpha$
(values taken from Wang, Lu \& Zhou 1998 and Constantin \& Shields
2003). Likewise, no correlation is found with the H$\beta$/C{\sc iv} ratio,
which can be used as a ionisation diagnostic, or with the H$\beta$ FWHM (obtained from Dewangan
2002; Kaspi \etal 2000) or EW (Marziani \etal 2003; V{\'
e}ron-Cetty, V{\' e}ron \& Gon{\c c}alves 2001); this is shown in Figure~\ref{hbetafwhm}. H$\beta$ line fluxes for the H$\beta$/C{\sc iv} ratio were
taken from Marziani \etal (2003), Cruz-Gonz{\' a}lez \etal (1994) and
Mulchaey \etal (1994). 

Comparing the slope of the Baldwin relation is problematic, as
different optical/UV lines show different slopes. For example, using the same cosmology as in this paper, Croom
\etal (2002) find a correlation for the C{\sc iv} line, produced in
the BLR, of EW~$\propto$~L$^{-0.128 \pm 0.015}$.  They also find
EW~$\propto$~L$^{-0.07 \pm 0.008}$ for the C{\sc iii}] blend (C{\sc
iii}], Al{\sc iii} and Si{\sc iii}]). Dietrich \etal (2002) give a
list of Baldwin effect slopes for different emission lines, varying
between $\sim$~$-$0.24 and +0.02, for H$_0$~=~65~km~s$^{-1}$~Mpc$^{-1}$ and $\Lambda$~=~0. They mention that including a cosmological constant of 0.7 changes the luminosities by $<$~20~per~cent. Green et al. (2001) find a
similar range, using H$_0$~=~50~km~s$^{-1}$~Mpc$^{-1}$ and $\Lambda$~=~0. The overall impression in the optical/UV is of a
correlation between the slope of the Baldwin relation and the ionisation potential, suggesting a change
in the average continuum shape with luminosity or redshift (depending
which dominates the correlation found in a particular study). It is
possible that the X-ray Baldwin effect found here is produced by a
corresponding variation in the shape of the continuum illuminating the fluorescing
material, although one might expect then to see some correlations with
the observed optical/UV lines.

The range of luminosities of the AGN in this sample over-laps with those investigated by Dietrich \etal (2002). Upon comparison of all the objects with those with log$\lambda$L$_{\lambda}$(1450\AA)~$>$~44, they find little change in the slope of the Baldwin relation for the different ranges, so the fact that QSOs tend to have larger L$_{UV}$/L$_{X}$ ratios than Seyfert galaxies is not likely to be an important factor.

\begin{figure}
\begin{center}
\includegraphics[clip,width=0.7\linewidth,angle=-90]{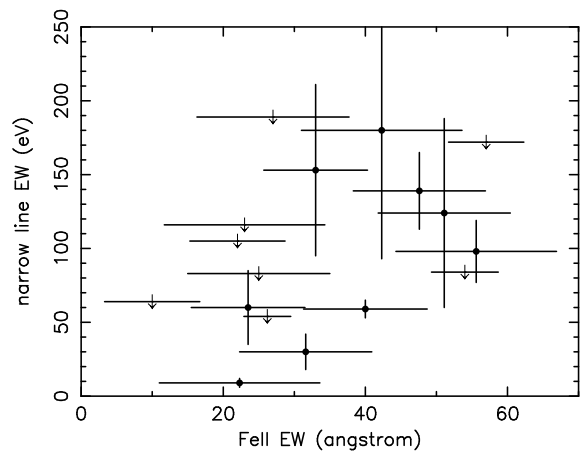}
\caption{The equivalent widths of the optical Fe~{\sc ii} and  
X-ray Fe~K$\alpha$ lines.}
\label{feii}
\end{center}
\end{figure}

Alternatively it could be argued that the iron abundance has been
underestimated and varies with luminosity.  Predicting the trend of
metallicity in AGN is complicated since their evolutionary
history is not known sufficiently well. BLR metallicities appear to be near-Solar,
but comparison of BLR observations with detailed photoionisation
models suggest that BLR metallicity does increase with luminosity
(e.g., Hamann \& Ferland 1993; Shemmer \& Netzer 2002). This is
oppposite to the trend expected if the X-ray
Baldwin effect is to be explained by an abundance effect. Deriving the iron abundance
from optical line ratios is made difficult due to the thermostatic
effect of Fe~{\sc ii} (which is a major coolant). For the present sample,
there is no correlation between the strength of the X-ray Fe~K$\alpha$
line and the optical Fe~{\sc ii} feature at $\lambda$4590 (Figure~\ref{feii}; Fe~{\sc ii} EWs from Marziani
\etal 2003).

The predicted EW is linearly dependent on column density in the above
equation, but Yaqoob et al. (2001) note that the approximations are
only valid for N$_{H}$~$<$~5.6~$\times$~10$^{23}$~cm$^{-2}$ (i.e.
assuming an Fe K absorption optical depth much less than unity).
Although some BLR models suggest higher column densities (e.g.,
Radovich \& Rafanelli 1994; Recondo-Gonzalez \etal 1997), values $\sim
10^{23} - 10^{24}$ cm$^{-2}$ are thought `normal'. 



Typical BLR velocities are of the order 5000 km~s$^{-1}$, which is
resolvable by {\it Chandra}. To date, the velocity width of the narrow
iron line has only been measured a few times, and the results are not entirely
consistent. Kaspi \etal (2002) resolved the line profile of NGC~3783
with the {\it Chandra} High Energy Transmission Grating Spectrometer
(HETGS), finding a FWHM velocity of $\sim$~1700~km~s$^{-1}$, which is
low for a BLR but consistent with originating in the inner part of the
torus. Likewise, Ogle \etal (2000) obtain a FWHM value of 1800~$\pm$~200~km~s$^{-1}$ for the narrow Fe~K$\alpha$ line detected in NGC~4151. The width of the iron lines in MCG$-$6$-$30$-$15 have also been measured with the HETGS. The situation is complicated by the presence of the various different components of the line in this object, both resolved and unresolved, narrow and broad. However, Lee \etal (2002) obtain a width of 11 000$^{+4600}_{-4700}$~km~s$^{-1}$ for the resolved narrow component when considering the full observation; this drops to 3600$^{+3300}_{-2000}$~km~s$^{-1}$ for the `high' flux state. Within the errors, these measurements are consistent with a FWHM of $\sim$~6000~km~s$^{-1}$, which would indicate that the line is not formed in the torus. Yaqoob \etal (2001) measure a FWHM of $\sim$~4500~km~s$^{-1}$ for NGC~5548, which also supports an origin in the BLR. 


It should be noted that, although the measurements are all broadly consistent, and indicate an origin in either the outer BLR or torus, there are relatively few grating measurements of the line, so it is unclear how narrow the line core truly is. An alternative possibility is that the `narrow' lines may actually be formed in the accretion disc itself. Nayakshin (2000a,b) discusses how a Baldwin effect could be produced by the ionisation of the skin of the disc, as the luminosity increases. A recent paper by Yaqoob et al. (2003) suggests that such narrow lines may come from a disc which is viewed face-on and has a flat emissivity profile.
Similar suggestions have been
made for optical/UV BLR lines. However, as mentioned above, no correlation is found here between the narrow iron line EW and the H$\beta$ parameters. There is a difference
between the BLRs found in Narrow Line and Broad Line Seyfert galaxies
-- the NLS1s show much narrower emission lines, hence their name. It
has been previously noted that NLS1s tend to show weaker H$\beta$
emission (e.g., Gaskell 1985) and this was confirmed upon comparing
the H$\beta$ EW of samples of NLS1s and BLS1s (taken from Marziani
\etal 2003) -- Student's T-test gave a very low probability of the two
groups originating in the same population, with the NLS1s having a
much lower mean EW. Considering the few objects in the present sample,
this difference in line strengths is also found. When comparing the
Fe~K$\alpha$ EWs, however, there is no appreciable difference between
the narrow- and broad-line samples, shown both by the T-test and {\sc
asurv} (to account for the upper limits). 

Possible support for the accretion disc origin lies in the variability of the narrow line in Mrk~841 (Petrucci \etal 2002). They find that the line varied between two observations separated by 15 hours, although, within the errors, the change in line flux is not large. Such a rapid variation is hard to explain if the line is formed in distant reprocessing matter, indicating its origin might be closer to the central engine; the line width is only constrained to be $<$~170~eV. To date, Mrk~841 is the only object in which this variation has been observed.

Overall, while the BLR remains a possible source for some of the narrow iron emission, we are unable to explain how the BLR alone could
produce the strongest observed narrow lines nor why the Baldwin
relation should exist. In particular, it is unclear why the most
luminous objects should have fairly normal BLRs compared to the
lower-luminosity objects, as judged from their optical/UV BLR lines,
yet have no detectable narrow 6.4 keV iron line.

\subsection{The molecular torus}
\label{torus}

Krolik, Madau \& {\. Z}ycki (1994) discuss the production of the iron
K$\alpha$ line in the torus. They compute that for a Thomson optical
depth, $\tau_{T}$, of 0.5--1, the EW of the line should be of the
order 100~eV, assuming an unobscured view; for $\tau_{T}$~=~2, this
value approximately halves, to 55~eV. They also find a small decrease
in EW for increasing opening angle of the torus. The value of
$\sim$~100~eV, for a low optical depth, is, as they point out, in
broad agreement with the EW measured in many Seyfert 1 galaxies,
suggesting that the Fe~K$\alpha$ emission line could be due solely to
reflection/fluorescence from the torus and not linked to the accretion
disc or BLR at all. Ghisellini, Haardt \& Matt (1994) also mention
that the torus could produce an emission line of EW~$\sim$~90~eV, if
the column density is $\go$~10$^{24}$~cm$^{-2}$. It should be noted that both of these papers used a value of 3.31~$\times$~10$^{-5}$  for the Solar abundance of iron, rather than the 4.68~$\times$~10$^{-5}$ assumed by Yaqoob \etal (2000). Assuming the EW scales linearly with the abundance, the values from Krolik \etal (1994) and Ghisellini \etal (1994) should be scaled to $\sim$~141 or $\sim$~127~eV, respectively, to compare to Yaqoob \etal's result.

As mentioned earlier, the presence of a neutral reflection component would indicate the existence of cool, Compton thick material (Guilbert \& Rees 1988; Lightman \& White 1988; George \& Fabian 1991), such as the torus. Some of the spectra here show evidence for such reflection, although the values are generally poorly constrained. This is not entirely surprising though, since most of the objects are at fairly low redshift, meaning that the rest frame bands covered by {\it XMM} do not extend much above $\sim$~10~keV, while the Compton reflection hump is expected to peak around 30--50~keV; reflection components may not be detected in the present data, but that is not to say that they do not exist. {\it If}, however, reflection components are not found in the spectra, this does not necessarily rule out the torus as the origin of the narrow line. Instead, the material of the torus could be Compton thin; if this were the case, no strong reflection component would be expected. Matt, Guainazzi \& Maiolino (2003) show that, for a Compton thin torus with N$_{H}$~=~2~$\times$~10$^{23}$~cm$^{-2}$, a fluorescent iron line with EW~$\sim$~80~eV can be formed. This is, therefore, a further possibility for production of the narrow line.

K{\"o}nigl \& Kartje (1994) discuss a dusty disc-driven hydromagnetic
wind model for the torus, finding that, if
L$_{IR}$~$\go$~1.5~$\times$~10$^{42}$(M/10$^{7}\msun$)~erg~s$^{-1}$,
where L$_{IR}$ is the 2--100~$\mu$m infrared luminosity, then the
radiation pressure force could be expected to flatten the dust
distribution; this causes the opening angle of the torus to increase,
leading to a reduction in the covering factor. This relates to Krolik
\etal (1994), who found that the EW of the Fe K$\alpha$ line decreases
slightly with increasing opening angle. In this way, an increase in luminosity and subsequent decrease in covering factor could lead to a smaller EW. This result is similar to work done by Mushotzky \& Ferland (1984), who proposed a luminosity-dependent ionisation parameter and, hence, covering factor to explain the original Baldwin effect.

An alternative explanation, suggested by Ohsuga \& Umemura (2001), is
that relatively low-luminosity AGN are more likely to contain dusty
walls of gas, supported by radiation pressure from a circumnuclear
starburst; the stronger radiation pressure from the more luminous AGN
may prevent these from forming.


If the X-ray Baldwin effect is, indeed, caused by a decrease in the
covering factor of the torus, then this has implications for the
number of Type-2 QSOs. Many Seyfert 2 galaxies are known, but very
few obscured QSOs have yet been discovered (e.g., Derry \etal 2003 and
references therein). A luminosity-dependent drop in the covering
factor provides goes some way towards explaining the lack of high luminosity
obscured sources, although obscuration should still arise due to mergers and the birth of the QSOs.

\section{Conclusions}
\label{sec:conc}

There is an apparent X-ray Baldwin effect for the narrow, neutral iron
line observed in many AGN: as the 2--10 keV rest frame luminosity
increases, the equivalent width of the line drops. The reason for this
correlation is uncertain, but one possibility is a decline in the
covering factor of the putative molecular torus as the luminosity increases. This decrease in covering fraction could be due to increased radiation pressure flattening the torus, leading to an increased opening angle and, hence, smaller covering factor. Although the BLR remains a possibility for the origin of some of the narrow line, it is difficult to explain how the strongest lines could be formed, or what it is that leads to the Baldwin effect.

The negative correlation between EW of the line and the luminosity of the object is not solely due to a dilution effect by beaming
in the radio-loud sources, since the effect is also observed in the
radio-quiet objects alone.

Given the resolution of EPIC and the {\it Chandra} HETGS, it is not
currently possible to rule out the origin of the line being either the
BLR clouds or reflection off distant matter (i.e., the
torus). Future observations with calorimeter-based X-ray detectors (such as the X-ray Spectrometer on-board {\it Astro-E2}, or {\it XEUS}) will provide sufficient spectral resolution, of the order a few hundred km~s$^{-1}$, to be able to resolve this issue.


\section{ACKNOWLEDGMENTS}
The work in this paper is based on observations with {\it
XMM-Newton}, an ESA
science mission, with instruments and contributions directly funded by
ESA and NASA. The authors would like to thank the EPIC Consortium for all their work during the calibration phase, 
and the SOC and SSC teams for making the observation and analysis
possible. 
This research has made use of the NASA/IPAC Extragalactic
Database (NED), which is operated by the Jet Propulsion Laboratory,
California Institute of Technology, under contract with the National
Aeronautics and Space Administation.
Support from a PPARC studentship is
acknowledged by KLP.

\end{document}